\documentclass[a4paper]{article}
\RequirePackage{amsbsy}
\RequirePackage{amssymb}
\RequirePackage{amsmath}
\RequirePackage{mathrsfs}
\RequirePackage{bm}
\newtheorem{thr}{Theorem}
\newtheorem{dfn}{Definition}
\newtheorem{td}[thr]{Theorem-Definition}
\begin{document}
%%%%%%%%%%%%%%%%%%%%%%%%%%%%%%%%%%%%%%%%%%%%%%%%%%%%%%%%%%%%%%%%%%%%%%%%%%%%%%%%%%%%%%%%%%%%%%%%%%%
\title{\huge{\bf{On a Completely Antisymmetric\\ Cartan Torsion\\ Tensor}}}
\author{Luca Fabbri\\
\footnotesize (THEORY GROUP, I.N.F.N.~ and Physics Department of Bologna University)}
\date{}
%%%%%%%%%%%%%%%%%%%%%%%%%%%%%%%%%%%%%%%%%%%%%%%%%%%%%%%%%%%%%%%%%%%%%%%%%%%%%%%%%%%%%%%%%%%%%%%%%%%
\maketitle
%%%%%%%%%%%%%%%%%%%%%%%%%%%%%%%%%%%%%%%%%%%%%%%%%%%%%%%%%%%%%%%%%%%%%%%%%%%%%%%%%%%%%%%%%%%%%%%%%%%
%%%%%%%%%%%%%%%%%%%%%%%%%%%%%%%%%%%%%%%%%%%%%%%%%%%%%%%%%%%%%%%%%%%%%%%%%%%%%%%%%%%%%%%%%%%%%%%%%%%
\begin{abstract}
We discuss the general structure of metric geometries, and how metricity implies the complete antisymmetry of Cartan tensor; an application in the frame of Lie group theory is given. Interpretations of the completely antisymmetric torsion in physical models are reviewed.
\end{abstract}
%%%%%%%%%%%%%%%%%%%%%%%%%%%%%%%%%%%%%%%%%%%%%%%%%%%%%%%%%%%%%%%%%%%%%%%%%%%%%%%%%%%%%%%%%%%%%%%%%%%
\section*{Introduction}
Theory of General Relativity is built up on the idea that our description of natural phenomena must be generally covariant, meaning that, even if we need a frame to represent nature, this frame cannot endow our description with information not contained in nature itself; after the mathematical translation of this idea, we find that General Relativity is written in the language of tensors: any physical object is expressed by a tensor and properties of physical objects are expressed by tensorial equations, so to remove any dependence on the system of reference whatsoever.

After tensorial geometry is developed, it turns out that the space is endowed with a metric structure and with a differential one, the former being represented by the metric tensor $g$, the latter being defined through a connection $\Gamma$, and, up to this point, these two entities are the two fundamental ones in the description of the geometrical properties of the space we want to study.

Using the connection, it is possible to define, beside the covariant derivative $D$, a couple of very particular tensors, the Cartan tensor $Q$ and the Riemann tensor $G$; also, it is possible to calculate the covariant derivative applied to the metric tensor itself $Dg$.

All these quantities are tensors that can be zero, and, side by side, different geometries can be defined: geometries in which all the three tensors are \emph{a priori} different from zero are considered for example by Hehl, McCrea, Mielke and Ne'eman in \cite{h-mc-m-ne} and by McCrea in \cite{mc}, and in the references therein; situations in which $Q=0$ are considered by Poltorak in \cite{p}; cases in which both $Q=0$ and $G=0$ are considered by Nester and Yo in \cite{n-y}; the condition $Dg=0$ gives rise to geometries considered by de Sabbata and Sivaram in \cite{ds-s}, for a general review, and, in more details, they are considered by Shapiro, by Obukhov, by Arcos and Pereira, by Watanabe and Hayashi, by Capozziello, Lambiase and Stornaiolo, respectively in \cite{s}, \cite{o}, \cite{a-p}, \cite{w-h}, \cite{c-l-s}, and in the references therein; situations in which $Dg=0$ is accompanied by $G=0$ are considered by de Andrade, Barbosa and Pereira in \cite{da-b-p}, while cases in which the condition $Dg=0$ is followed by the assumption $Q=0$ are the very well known metric geometries considered almost everywhere, for example in the classic text \cite{e} by Einstein; finally, if all three tensors are zero, the geometry reduces to be the one of the flat Minkowskian space.

Within the framework of the geometries characterized by the condition $Dg=0$ with a non-vanishing Cartan tensor, Cartan tensor itself does not undergo to any constraint; instead, we will see that some constraints are actually achieved, and we will have a look at the consequences these constraints will have, reviewing a couple of the most important physical models.
%%%%%%%%%%%%%%%%%%%%%%%%%%%%%%%%%%%%%%%%%%%%%%%%%%%%%%%%%%%%%%%%%%%%%%%%%%%%%%%%%%%%%%%%%%%%%%%%%%%
%%%%%%%%%%%%%%%%%%%%%%%%%%%%%%%%%%%%%%%%%%%%%%%%%%%%%%%%%%%%%%%%%%%%%%%%%%%%%%%%%%%%%%%%%%%%%%%%%%%
\section{Relativistic Theories with a\\ Completely Antisymmetric Cartan Tensor}
\label{sec:1}
Given the metric tensor $g$, the most general connection that can be defined is decomposable as
\begin{eqnarray}
\Gamma^{\kappa}_{\alpha\omega}=\Lambda^{\kappa}_{\alpha\omega}-
L^{\kappa}_{\phantom{\kappa}\alpha\omega}+K^{\kappa}_{\phantom{\kappa}\alpha\omega}
\label{conndec}
\end{eqnarray}
where: the part
\begin{eqnarray}
\Lambda^{\kappa}_{\alpha\omega}=\frac{1}{2}g^{\kappa \rho}
\left(\partial_{\alpha}g_{\rho \omega}
+\partial_{\omega}g_{\rho \alpha}
-\partial_{\rho}g_{\alpha \omega}\right)
\label{lcconn}
\end{eqnarray}
is a connection that defines a covariant derivative $\nabla_{\alpha}$ such that $\nabla_{\alpha}g_{\mu\nu}\equiv0$ and that is symmetric in the two lower indices, while conversely, these two conditions are together verified only by this connection, called Levi-Civita connection; the part
\begin{eqnarray}
L^{\kappa}_{\phantom{\kappa}\alpha\omega}=\frac{1}{2}g^{\kappa \rho}
\left(D_{\alpha}g_{\rho \omega}+D_{\omega}g_{\rho \alpha}-D_{\rho}g_{\alpha \omega}\right)
\label{aux}
\end{eqnarray}
is a tensor that is symmetric in the two lower indices; finally 
\begin{eqnarray}
K^{\kappa}_{\phantom{\kappa}\alpha\omega}=\frac{1}{2}(Q^{\kappa}_{\phantom{\kappa}\alpha\omega}
+Q_{\alpha\omega}^{\phantom{\alpha\omega}\kappa}+Q_{\omega\alpha}^{\phantom{\omega\alpha}\kappa});
\label{auxil}
\end{eqnarray}
is a tensor that is antisymmetric in the first two indices, called contortion tensor (see Wasserman \cite{wa}).

It has been showed by Hehl and Kr\"{o}ner and by Hehl in \cite{h-k} and \cite{h} that it is reasonable to assume the condition $Dg=0$ to hold.
 
We give the following 
\begin{dfn}
When a general connection defines a covariant derivative that acts upon the metric satisfying the condition 
\begin{eqnarray}
D_{\alpha}g_{\mu\nu}=0,
\label{metricity}
\end{eqnarray} 
called Metric-compatibility condition, or Metricity condition, the connection is called Metric Connection, and a geometry in which we have this condition is a Metric-compatible M Geometry.
\end{dfn}

After having assumed the metricity condition we have that the decomposition of the connection (\ref{conndec}) reduces to 
\begin{eqnarray}
\Gamma^{\kappa}_{\alpha\omega}=
\Lambda^{\kappa}_{\alpha\omega}+K^{\kappa}_{\phantom{\kappa}\alpha\omega};
\end{eqnarray}
explicitly, we can write it in the form
\begin{eqnarray}
\Gamma^{\kappa}_{\alpha\omega}=\Lambda^{\kappa}_{\alpha\omega}+
\left(\frac{Q_{\alpha\omega}^{\phantom{\alpha\omega}\kappa}+
Q_{\omega\alpha}^{\phantom{\omega\alpha}\kappa}}{2}\right)
+\frac{1}{2}Q^{\kappa}_{\phantom{\kappa}\alpha\omega}
\label{conndecmetr}
\end{eqnarray}
where the first term is the Levi-Civita connection and it is symmetric in the two lower indices, the second term is tensorial and symmetric in the same couple of indices and the third term is tensorial but antisymmetric in the same indices.

Now, we said that metric-compatible geometries are defined by the presence of metric connections, and, from a logical viewpoint, we have then two possibilities: (i) all the connections are metric, or (ii) some of the connections are metric, while some others are not; in order to distinguish them we will talk about, respectively: (i) metric-hypercompatible MH geometries, and (ii) proper metric-compatible pM geometries.

In the case (ii), we have that the most general form of the metric connections is given as in equation (\ref{conndecmetr}), while for the others the most general form is given as in equation (\ref{conndec}): but we see that the case in which $Dg$ is equal to zero is a particular case of a more general situation in which $Dg$ can assume any value, so that the former expression is a particular case of the latter expression, which turns out to be the most general type of connection we can have in case (ii); a pM space does \emph{not} give any additional information with respect to an M space, in which the most general connection is non-metric (while some connections happen to be metric anyway). Then, we can not consider (ii) to be a meaningful way to define metric-compatible geometries.

In order to have a meaningful definition of what a metric-compatible geometry should be, we have to define it as in case (i), to be a metric-hypercompatible MH geometry.

We have then justified the following 
\begin{dfn}
When all the connections define covariant derivatives that act upon the metric satisfying the conditions 
\begin{eqnarray}
D_{\alpha}g_{\mu\nu}=0,
\label{metricities}
\end{eqnarray}
now called Metric-Hypercompatibility conditions, or Metricity conditions, the connections are called Metric Connections, and a geometry in which we have these conditions is a Metric-Hypercompatible MH Geometry.
\end{dfn}

After having assumed the metricity condition for all the connections definable in a given space, we have that the decomposition of the most general connection is given as in 
\begin{thr}
In a metric-hypercompatible geometry, the most general connection is decomposable in the form
\begin{eqnarray}
\Gamma^{\kappa}_{\alpha\omega}=\Lambda^{\kappa}_{\alpha\omega}
+\frac{1}{2}Q^{\kappa}_{\phantom{\kappa}\alpha\omega}
\label{conndecmetrhyperfundamental}
\end{eqnarray}
where the first term is the Levi-Civita connection and the second term is the completely antisymmetric Cartan tensor of that connection.\\
$\square$
\footnotesize
\textit{Proof.} Let us consider the quantity
\begin{eqnarray}
\nonumber
\widetilde{\Gamma}^{\kappa}_{\alpha\omega}=\Lambda^{\kappa}_{\alpha\omega}+
\left(\frac{Q_{\alpha\omega}^{\phantom{\alpha\omega}\kappa}+
Q_{\omega\alpha}^{\phantom{\omega\alpha}\kappa}}{2}\right):
\end{eqnarray}
it is clear that it is a connection; since in a metric-hypercompatible geometry all the connections have to be metric, this is a metric connection; further, this connection is symmetric in the two lower indices: a symmetric-metric connection is necessary the Levi-Civita connection, so
\begin{eqnarray}
\nonumber
\Lambda^{\kappa}_{\alpha\omega}+
\left(\frac{Q_{\alpha\omega}^{\phantom{\alpha\omega}\kappa}+
Q_{\omega\alpha}^{\phantom{\omega\alpha}\kappa}}{2}\right)=
\widetilde{\Gamma}^{\kappa}_{\alpha\omega}\equiv\Lambda^{\kappa}_{\alpha\omega},
\end{eqnarray}
which gives the tensorial condition
\begin{eqnarray}
\nonumber
Q_{\alpha\omega}^{\phantom{\alpha\omega}\kappa}+
Q_{\omega\alpha}^{\phantom{\omega\alpha}\kappa}\equiv0
\end{eqnarray}
that expresses the antisymmetry in the first and second index of Cartan tensor; on the other side, Cartan tensor is by definition antisymmetric in the second and third index: and this gives the complete antisymmetry of Cartan tensor. Finally, considering the connection (\ref{conndecmetr}), it is immediate to see that, with a completely antisymmetric Cartan tensor, it reduces to 
\begin{eqnarray}
\nonumber
\Gamma^{\kappa}_{\alpha\omega}=\Lambda^{\kappa}_{\alpha\omega}
+\frac{1}{2}Q^{\kappa}_{\phantom{\kappa}\alpha\omega}
\end{eqnarray}
proving the theorem.
\normalsize
$\blacksquare$
\end{thr}

After this decomposition of the connection, the metric $g$ and the completely antisymmetric Cartan tensor $Q$ turn out to be the fundamental tensors of the tensorial calculus.

Riemann tensor $G$ can be written as
\begin{td}
In metric-hypercompatible geometries, we have that the most general Riemann tensor is decomposable in the form
\begin{eqnarray}
G^{\kappa}_{\phantom{\kappa}\alpha \sigma \mu}\equiv 
R^{\kappa}_{\phantom{\kappa}\alpha \sigma \mu}+
\frac{1}{2}(\nabla_{\sigma}Q^{\kappa}_{\phantom{\kappa}\alpha \mu}
-\nabla_{\mu}Q^{\kappa}_{\phantom{\kappa}\alpha \sigma})
+\frac{1}{4}(Q^{\kappa}_{\phantom{\kappa}\rho \sigma}Q^{\rho}_{\phantom{\rho}\alpha \mu}
-Q^{\kappa}_{\phantom{\kappa}\rho \mu}Q^{\rho}_{\phantom{\rho}\alpha \sigma}),
\end{eqnarray}
where the first term is the Riemann curvature tensor, written in terms of the Levi-Civita connection, that is in terms of the metric and the second term is written in terms of the completely antisymmetric Cartan tensor of the connection that defines the Riemann tensor; its unique independent contraction is $G^{\mu}_{\phantom{\mu} \alpha \mu \beta}=G_{\alpha \beta}$ called Ricci tensor, whose contraction is $G_{\alpha \beta}g^{\alpha \beta}=G$ called Ricci scalar.
\end{td}

\subsection{Lie Groups}
\label{sec:1.2}
Let us consider an application of the general structures studied above from the point of view of Lie theory of groups.

We have that 
\begin{thr}
In metric-hypercompatible geometries, if a continuous transformation has generators represented by a basis that admits a dual one, then they are Killing vectors; they define a connection whose Cartan tensor is equal to minus the structure coefficients of the anholonomic basis, and its Riemann tensor vanishes.\\
$\square$
\footnotesize
\textit{Proof.} Consider the basis of vectors $\{\xi_{(b)}^{\mu}\}$ together with the orthonormal dual basis $\{\xi^{(b)}_{\mu}\}$ such that $\xi^{(a)}_{\mu}\xi_{(b)}^{\mu}=\delta^{a}_{b}$ and $\xi^{(m)}_{\mu}\xi_{(m)}^{\nu}=\delta^{\nu}_{\mu}$; defining 
\begin{eqnarray}
\nonumber
\Gamma^{\alpha}_{\beta\gamma}=\xi_{(k)}^{\alpha}\partial_{\beta}\xi^{(k)}_{\gamma},
\end{eqnarray}
we see that it is a connection, and it has to be metric: so, we have that 
\begin{eqnarray}
\nonumber
0=\xi_{(a)}^{\alpha}0=\xi_{(a)}^{\alpha}D_{\alpha}g_{\rho\omega}=
\xi_{(a)}^{\alpha}(\partial_{\alpha} g_{\rho\omega}
-g_{\beta\omega}\xi_{(k)}^{\beta}\partial_{\rho}\xi^{(k)}_{\alpha}
-g_{\rho\beta}\xi_{(k)}^{\beta}\partial_{\omega}\xi^{(k)}_{\alpha})=\\
\nonumber
=\xi_{(a)}^{\alpha}\partial_{\alpha} g_{\rho\omega}
+g_{\beta\omega}\partial_{\rho}\xi_{(a)}^{\beta}
+g_{\rho\beta}\partial_{\omega}\xi_{(a)}^{\beta}\equiv L_{(a)}g_{\rho\omega},
\end{eqnarray}
that is, the Lie derivative of the metric vanishes, and so the vectors are Killing vectors.

It is a straightforward calculation to see that, given the previous connection, its Cartan tensor is equal to minus the structure coefficients. Finally, it is possible to prove by a direct calculation that its Riemann tensor vanishes.
\normalsize
$\blacksquare$
\end{thr}
%%%%%%%%%%%%%%%%%%%%%%%%%%%%%%%%%%%%%%%%%%%%%%%%%%%%%%%%%%%%%%%%%%%%%%%%%%%%%%%%%%%%%%%%%%%%%%%%%%%
%%%%%%%%%%%%%%%%%%%%%%%%%%%%%%%%%%%%%%%%%%%%%%%%%%%%%%%%%%%%%%%%%%%%%%%%%%%%%%%%%%%%%%%%%%%%%%%%%%%
%%%%%%%%%%%%%%%%%%%%%%%%%%%%%%%%%%%%%%%%%%%%%%%%%%%%%%%%%%%%%%%%%%%%%%%%%%%%%%%%%%%%%%%%%%%%%%%%%%%
\section{Physical Models with a\\ Completely Antisymmetric Torsion Tensor}
\label{sec:2}
After the decomposition of the connection, we have seen that the metric $g$ and the completely antisymmetric Cartan tensor $Q$ are the fundamental objects of metric-hypercompatible geometries.

When considered under the point of view of a physical quantity, Cartan tensor $Q$ is usually called Torsion.

We will now review some physical models that use a completely antisymmetric torsion tensor as a dynamical field.

\begin{itemize}
\item[\textbf{I}] \textbf{Higher-dimensional Theories.} As we have seen in Sect.~\ref{sec:1.2}, a Lie group admits a connection whose Riemann tensor vanishes; this is referred to the Flattening of a space as described by that connection, and in this case, the space is said to be Parallelizable: hence, we can say that it is possible to flatten a space that has the structure of a Lie group.

Nonetheless, there are spaces which are not Lie groups, but they are parallelizable in this sense: this is the case of the $7$-sphere $S^{7}$, which, being isomorphic to the unitary Octonions, which are non-associative, is not a Lie group, but a suitable completely antisymmetric Cartan tensor gives the possibility to get a vanishing Riemann tensor, and thus it is parallelizable; moreover, $S^{7}$ is the only sphere that can be parallelized.

As showed by Englert in \cite{en}, the flattening of $S^{7}$ comes from the choice of the complete antisymmetry of Cartan tensor; on the background of the present discussion, the complete antisymmetry of Cartan tensor is not a choice anymore: it is the most general Cartan tensor we can have. Consequently, the connection used by Englert is not a particular connection chosen \emph{ad hoc}, but it is the most general connection that we can use to parallelize $S^{7}$.

The possibility to squash $S^{7}$ is essential in the framework of Kaluza-Klein Multidimensional Theories; in these theories, the space is considered \emph{a priori} as a generic $n$-dimensional space, and then the number of dimensions is fixed by using phenomenological considerations: according to Witten's observation that $11$ is the only dimension for which a space is big enough to contain $U(1)\times SU(2)\times SU(3)$ and small enough to allow supersymmetry (\cite{wi}), $11$-dimensional spaces are quite an attractive choice for the background of KK theories.

In $11$-dimensional KK theory, the parallelization of the $S^{7}$ is the fundamental process for the decomposition of the $11$-dimensional space into the product of the $7$-dimensional compactified $7$-sphere, and a remaining non-compactified $4$-dimensional Minkowskian space-time; after that the vacuum space configuration is structured in $M^{(1,3)} \times S^{7}$, the general form of the metric is fixed, and the known physical fields can take place in it (for a general introduction to KK theories, see the original works of Kaluza and Klein, among the others, in \cite{kk1} and also in \cite{kk2}).

The completely antisymmetric torsion in KK theories is taken as a completely antisymmetric potential for a correspondent completely antisymmetric strength; this strength is the superfield we need to induce the spontaneous compactification mechanism for $11$-dimensional KK theory of supergravity (Englert \cite{en}).

Other applications of the completely antisymmetric torsion tensor can be found in String and Superstring theories, as described by Agricola, Friedrich, Nagy and Puhle in \cite{a-f-n-p}, by Strominger in \cite{st} and by Gauntlett, Martelli and Waldram in \cite{g-m-w}; in particular, Wormholes have been studied by Hochberg and Visser in \cite{h-v}.

For whom is concerned by the superfield of the $11$-dimensional KK theories, and finds arbitrariness in the choice of the number of dimensions, $4$-dimensional spacetime is then the only space gravitation can take place in.

\item[\textbf{II}] \textbf{Quadridimensional Theories.} Considering torsion and the Riemann tensor, it is possible to see that they verify the so-called Jacobi-Bianchi identities; when fully contracted, these identities get the form 
\begin{eqnarray}
\nabla_{\kappa}Q^{\kappa}_{\phantom{\kappa}\nu \mu}\equiv G_{\nu \mu}-G_{\mu \nu},
\label{j1}
\end{eqnarray}
which tells us that Ricci tensor has the antisymmetric part equal to the divergence calculated with respect to the Levi-Civita connection of the Cartan tensor, and 
\begin{eqnarray}
D_{\mu}G^{\mu}_{\phantom{\mu} \rho}-\frac{1}{2}D_{\rho}G
-G^{\mu \beta}Q_{\beta \mu \rho}+\frac{1}{2}G_{\mu \kappa \beta \rho}Q^{\beta \mu \kappa}\equiv0
\label{j2}
\end{eqnarray}
for Riemann tensor.

Now, considering the theory of matter fields, we know that their spin $S_{\alpha\rho\omega}$ and energy-momentum $T_{\alpha\omega}$ are tensors that are related by the coupled relationships
\begin{eqnarray}
D_{\mu}T^{\mu\nu}=-T_{\alpha\rho}Q^{\alpha\rho\nu}
+S_{\alpha\rho\sigma}G^{\alpha\rho\sigma\nu}
\label{E}
\end{eqnarray}
and
\begin{eqnarray}
\nabla_{\alpha}S^{\alpha\mu\nu}=-\frac{1}{2}(T^{\mu\nu}-T^{\nu\mu}),
\label{S}
\end{eqnarray}
as discussed in \cite{w-h} and \cite{h-vdh-k-n}, in which, using two complementary methods, the authors get the same result.

Coming back to the Jacobi-Bianchi fully contracted identities, it is obvious how the two sets of equations (\ref{j1})-(\ref{j2}) and (\ref{E})-(\ref{S}) look alike: this analogy is used to suggest the form of field equations of the theory to be
\begin{eqnarray}
Q^{\kappa\nu\rho}=kS^{\kappa\nu\rho}
\label{spin}
\end{eqnarray}
and \begin{eqnarray}
G^{\mu\rho}-\frac{1}{2}g^{\mu\rho}G=-\frac{k}{2}T^{\mu\rho}
\label{G}
\end{eqnarray}
in terms of the coupling constant $k$; then, once these field equations are given, the two sets of equations (\ref{j1})-(\ref{j2}) and (\ref{E})-(\ref{S}) do coincide.

Equations (\ref{spin})-(\ref{G}) are called Einstein-Sciama-Kibble field equations; and the theory they define is the Einstein-Sciama-Kibble theory: according to the picture drawn by the ESK theory, the spin of matter fields is the torsion, described by Cartan tensor $Q$, while the energy-momentum of mater fields acts geometrically by changing the metric of the spacetime $g$ (for a general discussion about ESK theory see, for example, the original papers by Einstein \cite{ein} and by Kibble and Sciama \cite{k} and \cite{sc}).

Watching at the metric as related, through the energy-momentum, to mass and torsion as related to spin, it is easy to recognize the analogies between the metric and torsion as fundamental quantities in Relativity and mass and spin as the fundamental quantum numbers that label elementary particles in terms of unitary irreducible representations of the Poincar\'{e} group. 

As discussed by Wigner in \cite{wig}, according to this classification, no constraint affects the mass, while spin is a number that is quantized, and whose values can only be of the form $\frac{k}{2}$ with $k \in \mathbb{N}$; the quantum number that labels spin can thus provide a classification of quantum matter fields.

Now, in terms of this classification of fields, we have that for spin-$0$ fields the spin tensor vanishes, for spin-$\frac{1}{2}$ fields it is completely antisymmetric, and for any other case in general it is non-completely antisymmetric, as discussed by Rarita and Schwinger in \cite{r-s}.

Thus said, it is clear how only fundamental fields of matter whose spin is equal to $0$ or $\frac{1}{2}$ can find place in this geometry, that is scalar fields and Dirac fields are the sole fundamental matter fields we can consider in our physical description of nature, according to the ESK theory.

Considering spin-$\frac{1}{2}$ fields, it is well-known how the Dirac-Fock-Ivanenko fermionic fields find a natural place in the ESK theory; the resulting Dirac-Fock-Ivanenko field equation is non-linear, and an autointeracting term arises (see, for example, \cite{h-d}): this new autointeracting term can provide the mechanism for CP violation in a spontaneous way as explained by Andrianov and Soldati in \cite{a-s} and by Andrianov, Giacconi and Soldati in \cite{a-g-s1} and \cite{a-g-s2}. Further, Chiral Anomalies have been treated in \cite{m} by Mielke and in \cite{k-m} by Kreimer and Mielke.

As the Dirac-Fock-Ivanenko field equation describes the behaviour of fundamental matter fields, the macroscopic approximation of this matter field equation has to reduce to Newton's equation of motion.

Given the line element $ds^{2}=g_{\mu\nu}dx^{\mu}dx^{\nu}$ with which we can build up the $4$-velocity $dx^{\mu}/ds=u^{\mu}$, we have that the equation of motion for a test body of mass $m$ reads
\begin{eqnarray}
\nonumber
mu^{\mu}D_{\mu}u^{\alpha}=F^{\alpha},
\label{eqmotion}
\end{eqnarray} 
where $F^{\alpha}$ is the covariant force acting on it; because in the framework of ESK theory the action of the gravitational field is already contained in the metric-symmetric part of the connection, the equation of motion in a gravitational field is a free equation of motion, so that it gets the form
\begin{eqnarray}
\nonumber
u^{\mu}D_{\mu}u^{\alpha}=0
\end{eqnarray}
called Autoparallel Equation, which represents the straightest line between two points in a space of given connection; because of the complete antisymmetry of Cartan tensor, we have that the latter equation reduces to 
\begin{eqnarray}
\nonumber
u^{\mu}\nabla_{\mu}u^{\alpha}=0
\label{geodesic}
\end{eqnarray}
called Geodesics Equation, which represents the shortest line between two points in a space of given metric, and it is the equation we would have had in absence of torsion.

This tells us that, even in presence of torsion, the geodesic equation is not distinguishable from the autoparallel equation, and this allows us to solve the AG paradox discussed by Fiziev in \cite{f}.

The fact that the autoparallel equation is not distinguishable from the geodesic equation is equivalent to the fact that torsion has no influence in the motion of macroscopic test bodies, and since torsion is spin, this means that spin does not affect the motion of test bodies in macroscopic situations; this is not surprising, for spin is a quantum effect, and it naturally disappears at macroscopic scales.

Anyway, although we can not detect torsion at macroscopic scales, we can detect it, almost paradoxically, at cosmological scales, where the presence of torsion filling up the whole universe in the early epoch could have had effects still measurable nowadays.

In the work \cite{g-mh}, G\"{o}nner and M\"{u}ller-Hoissen build up a cosmological model in which torsion is present as well as curvature in the generalized Friedman equations, and the most general torsion has only two independent components, namely
\begin{eqnarray}
\nonumber
Q_{jj0}=h(t)\\
\nonumber
Q_{ijk}=f(t)
\end{eqnarray}
where $t$ is the time labeled by $0$, and where $i,j,k$ are the spatial coordinates; in \cite{b}, B\"{o}hmer describes the particular situation in which the field $h$ is equal to zero: within the framework of our treatment, the completely antisymmetric torsion constrains the field $h$ to be zero, and the model considered by B\"{o}hmer is not one of the possible cases, but the only physical case this cosmological model can have. In this optic, B\"{o}hmer's \textit{ansatz} of exponential expansion of a universe in which torsion is the leading contribution for field equations can explain the inflation era without the introduction of other particles, beside the fact that it can explain why torsion nowadays might be a small but non-vanishing field we can actually detect by cosmological measures (see de Andrade in \cite{da}).

So, even if torsion is too small a field to be detected in a direct way, its effects on the evolution of the universe might be measured at cosmological scales, as discussed in \cite{ds-s} by de Sabbata and Sivaram; also, for a general discussion about macro and micro-gravity, see Hehl in \cite{he}.

Also, see the works \cite{bp-hn-s} and \cite{he} by de Berredo-Peixoto, Helayel-Neto and Shapiro and by Hehl for general considerations about matter fields in the ESK theory, through the so-called gauge theory for the Lorentz-Poincar\'{e} group.

Finally, in order to study Topological Invariants, Drechsler in \cite{d} proposed a new set of field equations that differs from that of the ESK theory, although always in a $4$ dimensional spacetime, and with a completely antisymmetric torsion.

Everything we discussed here refers to $4$-dimensional spacetime; anyhow, it is possible to consider also lower-dimensional spaces.

\item[\textbf{III}] \textbf{Lower-dimensional Theories.} In this case, the dimension can only be equal to $3$ or $2$, and accordingly:
\begin{itemize}
\item[\textbf{i}] \textbf{$3$-dimensional Theories.} For these theories, a completely antisymmetric torsion is proportional to the completely antisymmetric Levi-Civita tensor 
\begin{eqnarray}
\nonumber
Q_{ijk}=\phi \varepsilon_{ijk}
\end{eqnarray}
for a given pseudo-scalar field; models for this space have been proposed especially by Mielke and Baekler in \cite{m-b} and by Baekler, Mielke and Hehl in \cite{b-m-h}, as discussed in general by Blagojevic and Cvetkovic in \cite{b-c}.
\item[\textbf{ii}] \textbf{$2$-dimensional Theories.} In the case of $2$-dimensions, we have that the completely antisymmetric torsion always vanishes, making this latter case trivial.
\end{itemize}
\end{itemize}

This concludes the overview of some of the most fundamental physical models that use a completely antisymmetric torsion tensor as a dynamical field.
%%%%%%%%%%%%%%%%%%%%%%%%%%%%%%%%%%%%%%%%%%%%%%%%%%%%%%%%%%%%%%%%%%%%%%%%%%%%%%%%%%%%%%%%%%%%%%%%%%%
%%%%%%%%%%%%%%%%%%%%%%%%%%%%%%%%%%%%%%%%%%%%%%%%%%%%%%%%%%%%%%%%%%%%%%%%%%%%%%%%%%%%%%%%%%%%%%%%%%%
\section*{Conclusions}
In the present paper, we have considered what we defined to be a metric-hypercompatible geometry, in which Cartan tensor is completely antisymmetric; it has been given an application in the case of the Lie group theory. In terms of physical interpretation, the complete antisymmetric torsion supplies the condition needed for plenty of applications in complementary models of physical theories: in $11$-dimensional KK theories the complete antisymmetry of Cartan tensor is what allows the flattening process that squashes the $7$-sphere for the spontaneous compactification mechanism, while in the ordinary $4$-dimensional spacetimes the completely antisymmetric torsion is the spin of matter fields in the scheme of the ESK theory, and also $3$- and $2$-dimensional spaces have been considered.
%%%%%%%%%%%%%%%%%%%%%%%%%%%%%%%%%%%%%%%%%%%%%%%%%%%%%%%%%%%%%%%%%%%%%%%%%%%%%%%%%%%%%%%%%%%%%%%%%%%

\

%%%%%%%%%%%%%%%%%%%%%%%%%%%%%%%%%%%%%%%%%%%%%%%%%%%%%%%%%%%%%%%%%%%%%%%%%%%%%%%%%%%%%%%%%%%%%%%%%%%
\noindent \textbf{Acknowledgments.} I would like to thank Christian G.~B\"{o}hmer for the interesting discussions we have had together. 
%%%%%%%%%%%%%%%%%%%%%%%%%%%%%%%%%%%%%%%%%%%%%%%%%%%%%%%%%%%%%%%%%%%%%%%%%%%%%%%%%%%%%%%%%%%%%%%%%%%

\

%%%%%%%%%%%%%%%%%%%%%%%%%%%%%%%%%%%%%%%%%%%%%%%%%%%%%%%%%%%%%%%%%%%%%%%%%%%%%%%%%%%%%%%%%%%%%%%%%%%

%%%%%%%%%%%%%%%%%%%%%%%%%%%%%%%%%%%%%%%%%%%%%%%%%%%%%%%%%%%%%%%%%%%%%%%%%%%%%%%%%%%%%%%%%%%%%%%%%%%

\begin{thebibliography}{50}
\bibitem{h-mc-m-ne}
Friedrich W.~Hehl, J.~Dermott McCrea, Eckehard W.~Mielke and Yuval Ne'eman,
\textit{Phys. Rept.} \textbf{258}, 1 (1995).
\bibitem{mc}
J.~Dermott McCrea,
\textit{Class. Quant. Grav.} \textbf{9}, 553 (1992).
\bibitem{p}
A.~Poltorak,
arXiv:gr-qc/0407060.
\bibitem{n-y}
J.~M.~Nester and H.~J.~Yo,
arXiv:gr-qc/9809049.
\bibitem{ds-s}
Venzo de Sabbata and C.~Sivaram,
``Spin and Torsion in Gravitation'', chapter I.
World Scientific, Singapore (1994).
\bibitem{s}
I.~L.~Shapiro,
\textit{Phys. Rept.} \textbf{357}, 113 (2002).
\bibitem{o}
Y.~N.~Obukhov,
\textit{Int. J. Geom. Meth. Mod. Phys.} \textbf{3}, 95 (2006).
\bibitem{a-p}
H.~I.~Arcos and J.~G.~Pereira,
\textit{Int. J. Mod. Phys.} \textbf{D13}, 2193 (2004).
\bibitem{w-h}
T.~Watanabe and M.~J.~Hayashi,
arXiv:gr-qc/0409029.
\bibitem{c-l-s}
S.~Capozziello, G.~Lambiase and C.~Stornaiolo,\\
\textit{Annalen Phys.} \textbf{10}, 713 (2001).
\bibitem{da-b-p}
V.~C.~de Andrade, A.~L.~Barbosa and J.~G.~Pereira,\\
\textit{Int. J. Mod. Phys.} \textbf{D14}, 1635 (2005).
\bibitem{e}
Albert Einstein,
``The Meaning of Relativity''.\\
Princeton University Press, Princeton-N.J. U.S.A. (2004).
\bibitem{wa}
Robert H.~Wasserman,
``Tensors and Manifolds'', chapters 16 and 17.\\
Oxford University Press, New York-N.Y. U.S.A. (1992).
\bibitem{h-k}
F.~Hehl and E.~Kr\"{o}ner,
\textit{Z. Physik} \textbf{187}, 478 (1965).
\bibitem{h}
F.~Hehl,
\textit{Abhandl. Braunschweiger Wiss. Ges.} \textbf{18}, 98 (1966).
\bibitem{en}
F.~Englert,
\textit{Phys. Lett.} \textbf{B119}, 339 (1982).
\bibitem{wi}
E.~Witten,
\textit{Nucl. Phys.} \textbf{B186}, 412 (1981).
\bibitem{kk1}
H.~C.~Lee,
``An Introduction to Kaluza-Klein Theories''.\\
World Scientific, Singapore (1984).
\bibitem{kk2}
T.~Appelquist, A.~Chodos and P.~G.~O.~Freund,\\
``Modern Kaluza-Klein Theories''.
Addison-Wesley, (1987).
\bibitem{a-f-n-p}
I.~Agricola, T.~Friedrich, P.~A.~Nagy and C.~Puhle,\\
\textit{Class. Quant. Grav.} \textbf{22}, 2569 (2005).
\bibitem{st}
A.~Strominger,
\textit{Nucl. Phys.} \textbf{B274}, 253 (1986).
\bibitem{g-m-w}
J.~P.~Gauntlett, D.~Martelli and D.~Waldram,\\
\textit{Phys. Rev.} \textbf{D69}, 86002 (2004).
\bibitem{h-v}
D.~Hochberg and M.~Visser,
\textit{Phys. Rev.} \textbf{D58}, 44021 (1998).
\bibitem{h-vdh-k-n}
F.~W.~Hehl, P.~von der Heyde, G.~D.~Kerlick and J.~M.~Nester,\\
\textit{Rev. Mod. Phys.} \textbf{48}, 393 (1976).
\bibitem{ein}
Albert Einstein,
\textit{Sitzungsber. Preuss. Akad. Wiss. Berlin} \textbf{48}, 844 (1915).
\bibitem{k}
T.~W.~B.~Kibble,
\textit{J. Math. Phys.} \textbf{2}, 212 (1961).
\bibitem{sc}
D.~W.~Sciama, ``Recent Developments in General Relativity'', pag. 415.\\
Pergamon and PWN, Oxford U.K. (1962).
\bibitem{wig} 
E.~P.~Wigner,
\textit{Annals Math.} \textbf{40}, 149 (1939).
\bibitem{r-s}
W.~Rarita and J.~S.~Schwinger,
\textit{Phys. Rev.} \textbf{60}, 61 (1941).
\bibitem{h-d}
F.~W.~Hehl and B.~K.~Datta,
\textit{J. Math. Phys.} \textbf{12}, 1334 (1971).
\bibitem{a-s}
A.~A.~Andrianov and R.~Soldati,
\textit{Phys. Lett.} \textbf{B435}, 449 (1998).
\bibitem{a-g-s1}
A.~A.~Andrianov, P.~Giacconi and R.~Soldati,
JHEP \textbf{0202}, 30 (2002).
\bibitem{a-g-s2}
A.~A.~Andrianov, P.~Giacconi and R.~Soldati,\\
\textit{Grav. Cosmol. Suppl.} \textbf{8N1}, 41 (2002).
\bibitem{m}
E.~W.~Mielke,
AIP \textit{Conf. Proc.} \textbf{857B}, 246 (2006).
\bibitem{k-m}
D.~Kreimer and E.~W.~Mielke,
\textit{Phys. Rev.} \textbf{D63}, 048501 (2001).
\bibitem{f}
P.~P.~Fiziev,
arXiv:gr-qc/9808006.
\bibitem{g-mh}
H.~F.~M.~G\"{o}nner and F.~M\"{u}ller-Hoissen,\\
\textit{Class. Quant. Grav.} \textbf{1}, 651 (1984).
\bibitem{b}
C.~G.~B\"{o}hmer,
\textit{Acta Phys. Polon.} \textbf{B36}, 2841 (2005).
\bibitem{da}
L.~C.~Garcia de Andrade,
\textit{Int. J. Mod. Phys.} \textbf{D8}, 725 (1999).
\bibitem{he}
F.~W.~Hehl,
\textit{Found. Phys.} \textbf{15}, 451 (1985).
\bibitem{bp-hn-s}
G.~de Berredo-Peixoto, J.~A.~Helayel-Neto and I.~L.~Shapiro,\\
JHEP \textbf{2}, 3 (2000).
\bibitem{d}
W.~Drechsler,
\textit{Gen. Rel. Grav.} \textbf{15}, 703 (1983).
\bibitem{m-b}
E.~W.~Mielke and P.~Baekler,
\textit{Phys. Lett.} \textbf{A156}, 399 (1991).
\bibitem{b-m-h}
P.~Baekler, E.~W.~Mielke and F.~W.~Hehl,
\textit{Nuovo Cim.} \textbf{B107}, 91 (1992).
\bibitem{b-c}
M.~Blagojevic and B.~Cvetkovic,
arXiv:gr-qc/0412134.
\end{thebibliography}
\end{document}